\documentclass[twocolumn,showpacs,floats,prc,preprintnumbers,amsmath,amssymb]{revtex4}
\usepackage{graphicx}
\usepackage{dcolumn}
\usepackage{bm}
\def\fslash#1{#1 \!\!\! \slash} 

\renewcommand{\(}{\left(}
\renewcommand{\)}{\right )}
\renewcommand{\[}{\left [}
\renewcommand{\]}{\right ]}

\def\beq{\begin{equation}}
\def\eeq{\end{equation}}
\def\pa{\partial}

\def\bea{\arraycolsep .1em \begin{eqnarray}}
\def\eea{\end{eqnarray}}

\def\vp{{\bf p}}

\def\vk{{\bf k}}

\def\Tr{{\rm Tr}}
\def\rnn{{\rho NN}}

\let\om=\omega

\let\no=\nonumber

\def\eq#1{(\ref{#1})}
\def\refr#1{\cite{#1}}

\def\s0#1#2{\mbox{\small{$ \frac{#1}{#2} $}}}
\def\0#1#2{\frac{#1}{#2}}

\def\anp#1#2#3{Adv.\ Nucl.\ Phys. \ {\bf #1}, #2 (#3)}
\def\plb#1#2#3{Phys. Lett. {\bf B #1}, #2 (#3)}
\def\npa#1#2#3{Nucl. Phys. {\bf A #1}, #2 (#3)}
\def\npb#1#2#3{Nucl. Phys. {\bf B #1}, #2 (#3)}
\def\prc#1#2#3{Phys. Rev.  {\bf C #1}, #2 (#3)}
\def\prd#1#2#3{Phys. Rev. {\bf D #1}, #2 (#3)}
\def\prl#1#2#3{Phys. Rev. Lett. {\bf #1}, #2 (#3)}
\def\ann#1#2#3{Ann. Phys. {\bf #1}, #2 (#3)}
\def\anp#1#2#3{Adv. Nucl. Phys. {\bf #1}, #2 (#3)}
\def\pr#1#2#3{Phys. Rep. {\bf #1}, #2 (#3)}
\def\epja#1#2#3{Eur.\ Phys.\ J.{\bf A #1}, #2 (#3)}

\def\ptp#1#2#3{Prog.\ Theor.\ Phys. \ {\bf #1}, #2 (#3)}

\def\ijmpe#1#2#3{Int.\ J.\ Mod.\ Phys.\ {\bf E #1}, #2 (#3)}
\def\jhep#1#2#3{J.\ High Energy Phys. {\bf #1}, #2 (#3)}
\def\ibid#1#2#3{{\it ibid.}, {\bf #1}, #2 (#3)}

\begin{document}

\title{{\bf Dispersion relation of the $\rho $ meson in hot and dense nuclear matter}
}
\author{Ji-sheng Chen$^{a}$~~~~~~
Jia-rong Li$^{b}$~~~~~~Peng-fei Zhuang$^{a}$}
\address{$^a$Physics Department, Tsinghua University, Beijing 100084, 
People's Republic of China\\
$^b$Institute of Particle Physics,
 Hua-Zhong Normal University,
Wuhan 430079, 
People's Republic of China}

\thispagestyle{empty}
\begin{abstract}
The dispersion relation of $\rho$ meson in both timelike and spacelike
regimes in hot and dense nuclear medium is analyzed and compared with $\sigma $ meson
based on the quantum hadrodynamics model. 
The pole and screening masses of $\rho $ and $\sigma$ are discussed. 
The behavior of screening mass of
$\rho$ is different from that of $\sigma$ due
to different Dirac- and Fermi-sea contributions at finite temperature and density.
\end{abstract}
\pacs{14.40.Cs, 11.55.Fv, 11.10.Wx}
\maketitle
\par 
Heavy-ion collision physics has stimulated intense investigations of
the properties of strongly interacting, hot and dense nuclear matter\refr{rapp2000}. 
Among the proposed signals for detecting quark-hadron phase
transition, dileptons and photons are considered to be the clearest ones because
they can penetrate the medium almost
undisturbed and reflect the property of the fireball formed in the
initial stage of collisions\refr{shuryak1978}.
Furthermore, the dileptons from the decay of light vector mesons 
can be considered as possible signals of the partial chiral symmetry
restoration. Especially, the property of $\rho$ in hot and dense
environment has attracted much attention in the literature due to its
relatively larger
decay width compared with
$\om $ and $\phi
$\refr{teodorescu2001,gale1991,hatsuda1992}.
It is interesting that the $\rho$ mass decreasing mechanism 
can be used to explain the low invariant mass dilepton enhancement in
central $A-A$ collisions
observed by CERES-NA45\refr{brown1991,agakichiev1995,ligq1995}.  

From the point of view of many-body theory,
the collective effect of medium on a meson
is reflected by its full propagator, which determines its dispersion relation
as well as the response to the external
source\refr{kapusta1989,chin1977,saito1989}. 
Due to the broken Lorentz symmetry, the dispersion relations for
longitudinal and transverse modes of vector mesons are different.
However, the timelike and spacelike regimes 
are related to each other through the dispersion relation as in
the case of QED\refr{rebhan2001}. 
With vector meson dominance model,
the $\rho $ meson screening mass is an important quantity 
related to the EM (electromagnetic) Debye mass and to the emissivity of dileptons and photons produced in
 heavy-ion collisions. For example, the screening mass in spacelike limit
is associated with the isospin fluctuations which can be used as a potential
signature of QGP (quark-gluon plasma) formation\refr{eletsky1993}. Furthermore, the scalar quark density
fluctuation of QCD is related to the spacelike limit
of in-medium self-energy of $\sigma$ as pointed out in Ref.\refr{chanfray2001},
where the contribution of free nucleons at $T=0$
is analyzed through
one-loop $NN^{-1}$ excitation. In Refs. \refr{shiomi1994,chen2002},
it was found that the Dirac-sea contribution to the pole mass of $\rho $ dominates over 
Fermi sea's. In this paper, we discuss the dispersion relations of $\rho$
and $\sigma$ in both spacelike
and timelike regimes determined by the pole positions of their in-medium propagators.
The medium effects on $\rho $ and $\sigma$ at finite temperature $T$ and baryon density $\rho _B$
are taken into account in the framework of quantum hadrodynamics model(QHD)
through the in-medium nucleon excitation.

\begin{figure}[ht]
\includegraphics[width=8.0 cm,angle=-0]{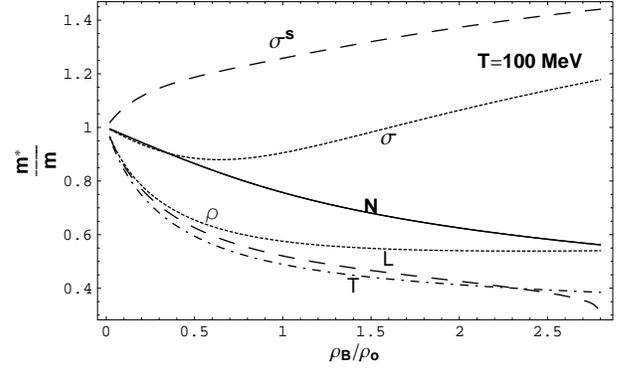}
\hfill\vskip0.1cm
\caption{
\small Effective masses 
as functions of scaled density at temperature $T=100$ MeV.
The solid line represents the effective nucleon mass(labeled as $N $).
The dotted lines are for pole masses($\rho $ and $\sigma$,
respectively),
dot-dashed for transverse screening mass ($T$). The  dashed lines indicate the
longitudinal screening mass($L$) of $\rho $ 
and screening mass($\sigma ^s$) of $\sigma$, respectively.}\label{fig1}
\end{figure}

We start from QHD-I to obtain the effective nucleon mass $M_N^*$ and
effective chemical potential $\mu ^*$ for discussing the in-medium meson property. 
In the relativistic Hartree
approximation, 
the self-consistent equations for $M_N^*$ and $\mu ^*$ can be written as
\refr{serot1986,chen2002}
\bea\label{selfconsistent}
&&M_N^*-M_N =-\0{g_\sigma^2}{m_\sigma^2}\0{4}{(2\pi )^3} \int d^3
\vp \0{M_N^*}{\om }\[n_B
+\bar{n}_B\]
\no\\&&~~~
+
\0{g_\sigma^2}{m_\sigma^2}\01{\pi ^2} \[M_N^{*3}\ln
\(\0{M_N^*}{M_N}\)-M_N^2 (M_N^*-M_N)
\right.\no\\
&&\left. 
~~~-\052 M_N (M_N^*-M_N)^2-\0{11}6
(M_N^*-M_N)^3\],\\
&&\mu ^* -\mu =-g_\om^2 \rho _B /m_\om^2,
\eea
where 
$\om =\sqrt{{M_N^*}^2+p^2}$ is the nucleon energy and
the baryon density $\rho _B $ is defined by
\bea\label{chemical}
\rho _{B} =\04{(2\pi )^3}\int d^3 \vp \[n_B -\bar{n}_B\],
\eea
with
$n_B$(${\bar n_B}$) being the Fermi-Dirac distribution functions for
(anti-)baryons, respectively.
The 
coupled equations can be solved
numerically 
with the parameters determined by fitting the binding energy at normal
nuclear density $\rho _0$ given in Table.\ref{para}. The $M_N^*$ decreases with increasing density
$\rho _B$ at fixed temperature(see Fig.\ref{fig1}), analogously to the result of mean field theory
neglecting the vacuum fluctuations\refr{chen2002}. 
The effective chemical potential $\mu ^*$ will affect the properties of
mesons indirectly through the distribution functions.

In Minkowski space, the polarization tensor $\Pi ^{\mu\nu}(k)$ of $\rho $
can be divided into two parts with the standard
projection
tensors $P_L
^{\mu\nu}$ and $ P^{\mu\nu }_T$ according to
\bea
\Pi ^{\mu\nu} (k) = \Pi _L (k) P_L ^{\mu\nu} +\Pi _T (k) P_T^{\mu\nu}
\eea
with 
\bea
\Pi_L(k)=\0{k^2}{\vk ^2 }\Pi ^{00}(k),~~~~~~~
\Pi_T(k)=\012 P^{ij}_T\Pi_{ij}(k).
\eea

With the effective Lagrangian for $\rho NN$ interactions\refr{machleidt1987}
 \bea {\cal L}_{\rho NN }=g_{\rho
NN} \({\bar \Psi} \gamma _\mu \tau ^a \Psi V_a ^\mu - \0{\kappa
_\rho}{2 M_N} {\bar \Psi}\sigma _{\mu\nu }\tau ^a\Psi \pa ^{\nu
}V_a^\mu \),\no \eea where $V_a^\mu$
is the $\rho$ meson field and $\Psi $ the nucleon field, the polarization
tensor is given in random phase approximation (RPA) by
\bea
&&\Pi ^{\mu\nu}(k) =2 g^2_{\rnn} T \sum _{p_0}
 \int \0{d^3  {\bf p}}{(2 \pi
)^3}\no\\
&&~~~~~\Tr \[\Gamma ^\mu(k) \0{1}{\fslash{p}-M_N^*}\Gamma ^\nu (-k)
\0{1}{(\fslash{p} -\fslash{k})-M_N^*} \],
\eea 
with $\Gamma ^\mu =\gamma ^\mu +({ik_\rho }/{2 M_N })\sigma^{\mu\nu} k_\nu $.
The temperature and effective chemical potential are hidden in 
the zero-component of nucleon momentum via ${p_0}
=(2n +1 )\pi T i +\mu ^*$. 
With the residue theorem, 
one can separate the
polarization tensor into two parts
\bea
\Pi ^{\mu\nu }(k) =\Pi ^{\mu\nu} _F (k) +\Pi ^{\mu\nu} _D (k),
\eea
where $\Pi ^{\mu\nu}_F(k)$ corresponds to the
particle-antiparticle contribution of the Dirac sea at $T=0$ and
$\Pi ^{\mu\nu}_D(k)$ arises from the particle-hole
contribution\refr{kurasawa1988,jean1994,dutt1997}.
The various components of $\Pi ^{\mu\nu}_D(k)$ are listed in the appendix
and $\Pi ^{\mu\nu}_F(k)$ can be found in Ref.\refr{chen2002}.

For vector meson excitation in the medium, the dispersion
relations determined by the pole positions of the full propagator
$D^{\mu\nu}$ for longitudinal and transverse branches are different,
while the pole masses determined by 
taking the limit $\Pi _{L,T}(k_0,|\vk|\rightarrow 0)$ for $L$ and $T$ modes are
consistent\refr{chen2002}. As pointed out in Ref.\refr{eletsky1993}, 
the screening mass determined by $\Pi _L (0,|\vk|\rightarrow 0) $
can be related to the isospin fluctuations.
In general case including the Dirac sea contribution and the vacuum mass
$m_\rho$,
the screening masses are defined by 
the pole positions of full propagators related with the finite momentum self-energy $\Pi (0,
\vk)$\refr{shiomi1994,rebhan2001}
\bea\label{dd}
\vk ^2+m_\rho ^2 +\Pi_{L,T} (0,\vk)=0.
\eea
The screening (Debye) masses $M_D=-i|\vk|$ are the inverse Debye screening lengths
and reflect the collective effects of medium, i.e.,
the damping characteristic $e^{-|\vk | x}$ of the excitations with purely imaginary wave
numbers. 
The self-consistent numerical results for $M_D$ determined by 
\eq{dd} are indicated in Fig.\ref{fig1},  where the effective nucleon
mass $M_N^*$ and the effective pole mass of $\rho$ are also shown for comparison.

It is interesting to discuss the in-medium property of scalar meson $\sigma$ with QHD 
and compare it with the vector mesons. 
At zero temperature, the property of $\sigma$ has
been discussed in Refs.
\refr{iwasaki00,caillon1993}. As pointed out in the introduction, the screening mass of $\sigma $ in spacelike limit 
 has been discussed 
at zero temperature by considering the one-loop $NN^{-1}$ excitation 
with free nucleon gas\refr{chanfray2001}.
At finite temperature, the $\sigma $                       
meson self-energy with RPA is
\bea
\Pi_\sigma (k)=2 g_\sigma ^2 T\sum _{p_0}\int \0{d^3{\bf p}}{(2\pi
)^3}
 \Tr \[\0{1}{\fslash{p}-M_N^*}
\0{1}{(\fslash{p} -\fslash{k})-M_N^*} \],\no
\eea
which can be reduced to
\bea
\label{si}
&&\Pi_ \sigma(k)=\frac{3 g_{\sigma}^2}{2 \pi^2} \left [3M_N^{*^2} -4 M^*_N
M_N+M_N^2 
\nonumber\right. \\&&\left.
-
(M_N^{*^2} - M_N^2) \int_{0}^{1}
\ln\frac{M_N^{*^2}
- x(1 - x)k^2}{M_N^2} dx \nonumber\right. \\
  &&\left.- \int_{0}^{1} (M_N^2 - x(1 - x)k^2) \ln\frac{M_N^{*^2} - x(1 - x)k^2}{
M_N^2 - x(1 - x)k^2} dx\right ]
\no\\&&
+\0{g_\sigma ^2}{\pi ^2}\int \0{p^2 dp }{\om
} (n_B+{\bar n_B})\[2+\0{k^2-4 {M^*_N}^2  }{4 p |\vk|}(a+b
)\],\eea
where
\bea
a=&&\ln\0{k^2-2 p |\vk|-2 k_0 \om }{k^2+2 p |\vk|-2 k_0 \om},~~~
b=\ln \0{k^2-2 p |\vk|+2 k_0 \om }{k^2+2 p |\vk|+2 k_0 \om}
\no
\eea
with $k^2=k_0^2-\vk ^2$.
It is neccesary to note again that here we discuss the full propagator
$D_\sigma$ with the in-medium nucleons. 
The effective masses of $\sigma $ meson defined
analogously to those of $\rho$ are also displayed in Fig.\ref{fig1}.

The pole masses $m^*_\rho $, $M_N^*$ and $m_\sigma ^*$ 
venus
$\rho _B$ at fixed $T$ behave very differently. 
The effective nucleon mass decreases monotonously with increasing density,
while the effective pole mass of $\rho $ decreases at first and then becomes
saturated. The pole mass
$m_\sigma ^*$ also decreases in the low density region.
As for the screening mass behavior, the
longitudinal and transverse Debye ones of $\rho $ decrease,
but the one of $\sigma $ increases with increasing density.
\begin{figure}[ht]
\includegraphics[width=8.0cm,angle=-0]{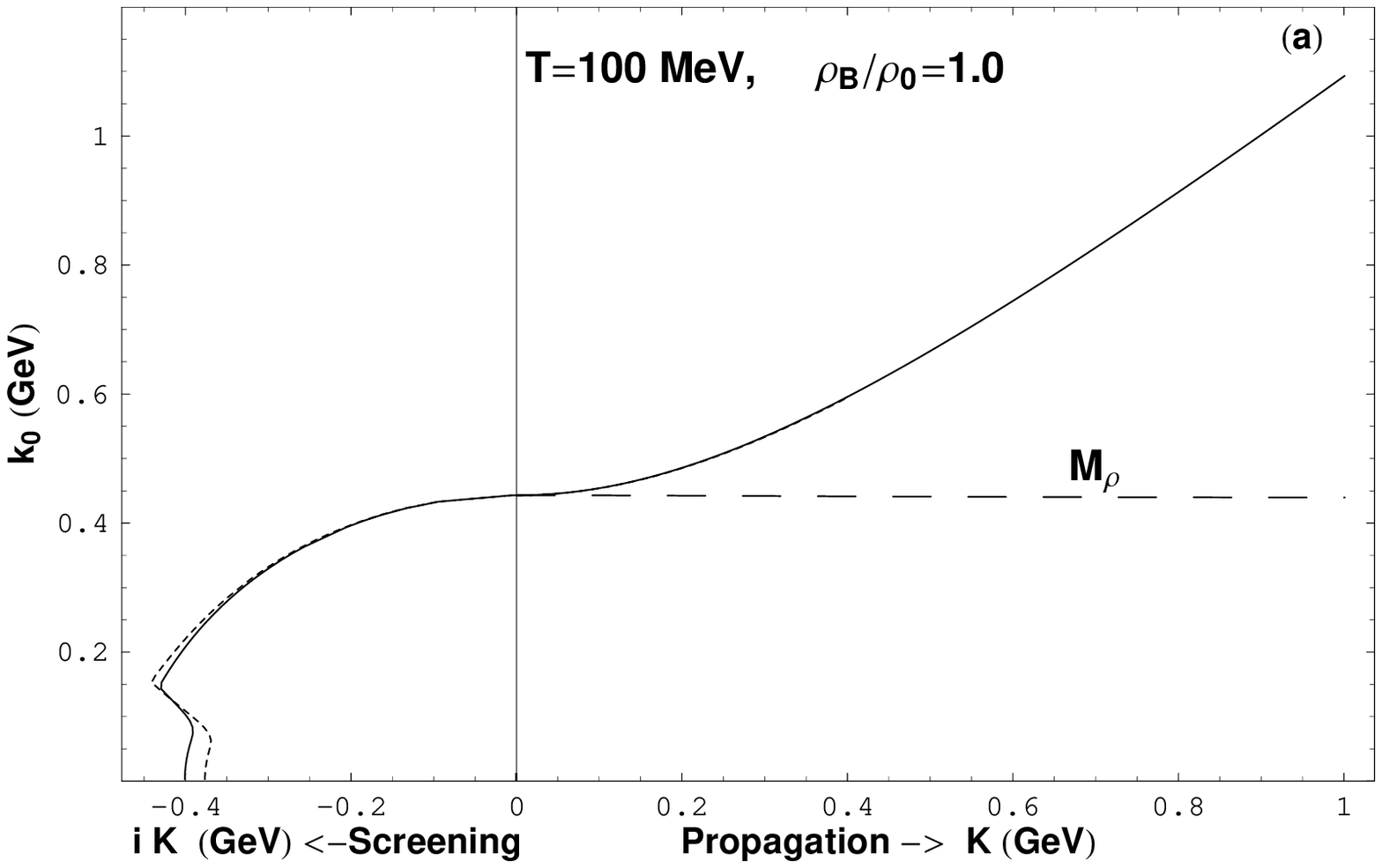}
\hfill\vskip0.2cm
\includegraphics[width=8.0cm,angle=-0]{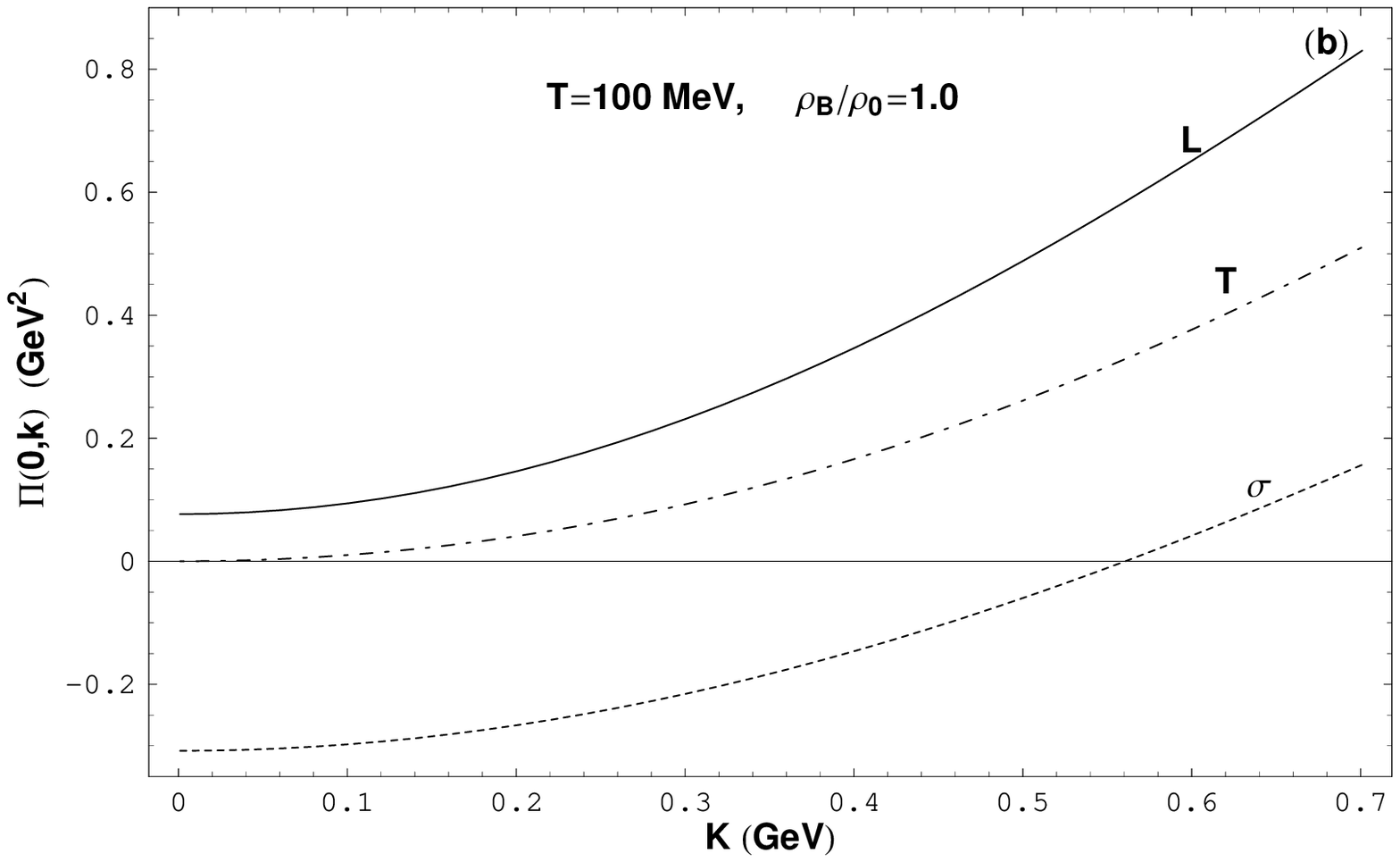}
\caption{
\small (a) Dispersion relation curves for $\rho $.  The solid line corresponds to
$L$
mode, dot-dashed to $T$ mode and long-dashed to invariant mass
$M_\rho=\sqrt{k_0^2-\vk ^2}$;
(b) $\Pi (0,\vk ) $ for $\rho $ and $\sigma$.
The solid line and dot-dashed lines correspond respectively to $L$ and $T$
modes
of $\rho $, and 
the dotted line is for $\sigma$.  
}\label{fig2}
\end{figure}

The corresponding
dispersion relation curves calculated from the pole position of the full
propagator $D^{\mu\nu}$ in both timelike and spacelike regions for $\rho $ are shown in
the upper panel of Fig.\ref{fig2}. 
Due to the tensor(magnetic) coupling and the relatively smaller coupling
constant $g_\rnn$ compared with $\om $ meson, the invariant in-medium mass $M_\rho =\sqrt {k_0^2-\vk ^2}$
is almost a
constant in the timelike region. The dispersion relation curves for the
longitudinal and transverse branches almost coincide
and can only be separated from each other in the spacelike
screening region.
For comparison, the spacelike tensors $\Pi (0,\vk)$ of $\rho$ and $\sigma$ are
shown in the lower panel of Fig.\ref{fig2}.
For $\rho$, in the limit $|\vk| \rightarrow 0$ the Dirac-sea and the
tensor coupling contributions vanish, and the Fermi sea contributes only to the
longitudinal mode. Therefore,  the screening mass determined by the
spacelike limit of $\Pi (0,\vk )$ will be very different from what we
showed in Fig.\ref{fig1}.
The difference between the $L$ and $T$ modes dominated by the Fermi-sea
contribution increases with increasing momentum  $K=|\vk|$. For $\sigma $,
$\Pi (0,\vk )$  contains both Dirac- and Fermi-sea
contributions in the spacelike limit. It is the Fermi-sea contribution
which leads to a negative
$\Pi (0,\vk) $ in the low $|\vk|$ region.  

In summary, we have analyzed the dispersion relations of $\rho $ and $\sigma $ in
hot and dense nuclear environment in the framework of QHD in 
timelike and
spacelike regimes. 
The pole and screening masses of $\rho $ are found to decrease with 
increasing density.
Although the pole mass of 
$\sigma$ decreases in the low density region, the screening mass behaves
very differently from those of $\rho$ at finite temperature and density.
This difference is attributed to the corresponding Dirac- and Fermi-sea contributions.
\vskip0.1cm
{\bf Acknowledgments} This work was supported
by the NSFC under Grant Nos 10135030, 10175026, 19925519 and the China Postdoc Research Fund.

\appendix
\begin{table}
\small  
 \caption{Parameters of QHD-I determined at normal nuclear density $\rho _0=0.1484fm
^{-3}$.  The masses are in (MeV). }
\begin{ruledtabular}     
\begin{tabular}{cdccccccc}
        &g_\sigma ^2& $g_\om ^2$ &$m_\sigma $& $m_\om $ &$g_\rnn $
&$ k_\rho $&$m_\rho$&$M_N$  \\ 
     \tableline
        &  54.289 & 102.770 &458& 783 & 2.63&6.1&770&939  \\
   \end{tabular}
 \label{para}
\end{ruledtabular}
\end{table}
\section{}
The ingredients of $\Pi ^{\mu\nu }_D$ with the similar expressions of $a$ and $b$
in \eq{si} are
\bea
&&\Pi^{00}_D(k)=\Pi ^{00}_{1D}+\Pi ^{00}_{2D}+\Pi ^{00}_{3D},\no\\
\Pi ^{00}_{1D}&&=-2  \(\0{g_\rnn  }{2 \pi }\)^2 \int \0{p^2 dp }{\om } (n_B
+{\bar n_B})
\no\\&&
\[4 + \0{k^2 - 4 \om k_0 + 4 \om ^2 } { 2 p |\vk| }a\no 
+ (\om \rightarrow -\om)
\],
\no\\
\Pi ^{00}_{2D}&&= 4 |\vk |  \(\0{g_\rnn }{2 \pi }\)^2 \0{k_\rho }{2 M_N
}M_N^*\int \0{p dp }{\om } (n_B +{\bar n_B}) (a + b),
\no\\
\Pi ^{00}_{3D}&&=2 \(\0{g_\rnn }{2 \pi }\)^2 \(\0{k_\rho }{2 M_N }\)^2 \int
\0{p^2 dp }{\om } (n_B+{\bar n_B})\no\\
&&\[ 
4 k^2 _0
+\0{\vk ^2 (k^2 - 4 p^2 )+(k^2 -2 k_0\om  )^2}{2 p |\vk|}a 
+(\om \rightarrow -\om )
\];\no\\
&&\Pi^{0i}_D(k)=\0{k^0k^i}{\vk ^2} \Pi ^{00}_D(k);\no\\
&&\Pi^{ij}_D(k)=(A_1+A_2+A_3)\delta ^{ij} 
+(B_1+B_2+B_3) \0{k^ik^j}{\vk ^2},\no\\
A_1&&=\(\0{g_\rnn  }{2 \pi }\)^2\int
\0{p^2 dp }{\om } (n_B +{\bar n_B})\no\\&&
\[\0{4 (\vk^2 +k_0^2)}{\vk ^2}
\right.\no\\
&&\left.
-
\0{\vk ^4 -k_0^2 (k_0-2 \om )^2 +4 \vk ^2 (p^2 -k_0 \om )}{2 p |\vk |^3 } a
-(\om \rightarrow -\om)
\],\no\\
A_2&&=\0{k^2}{\vk^2}\Pi ^{00}_{2D},\no\\
A_3&&=-k^2 \(\0{g_\rnn  }{2 \pi }\)^2 \(\0{k_\rho }{2 M_N }\)^2  \int
\0{p^2 dp }{\om } (n_B 
+{\bar n_B})
\[ \0{4k^2}{\vk ^2}+
\right.\no\\
&&\left.
\0{k ^4 +4 k^2 \om (\om -k_0) +4 \vk ^2 (p^2 -\om ^2 )}{2 p |\vk |^3 }
a
+(\om \rightarrow -\om)
\],\no\\
B_1&&=\(\0{g_\rnn }{2 \pi }\)^2 \int
\0{p^2 dp }{\om } (n_B+{\bar n_B})
\[- \0{4(\vk^2 +3 k_0^2)}{\vk ^2}
\right.\no\\
&&\left.+    
\0{\vk ^4 -3 k_0^2 (k_0-2 \om )^2+2 \vk ^2(k_0^2+2p^2-2 k_0\om )}{2 p |\vk |^3 }a
\right.\no\\
&&\left.+(\om \rightarrow -\om)
\],\no\\
B_2&&=\Pi ^{00}_{2D},\no
\label{a3}\eea
\begin{widetext}
\bea
B_3&&=\(\0{g_\rnn  }{2 \pi }\)^2 \(\0{k_\rho}{2 M_N}\)^2 \int
\0{p^2 dp }{\om } (n_B
+{\bar n_B})
\[\0{4(k_0^4 + k^4)}{\vk ^2}
\right.\no\\
&&\left.~
+\0{k^2 (2 k_0^4+k^4)+4k_0 k^2 (2 k_0^2+k^2)\om  -4 (2 \vk ^2 +k^2) \vk^2 p^2+4
(2k_0^4-k_0^2k^2+2k^4)\om^2 }{2 p |\vk
|^3 }
a
+(\om \rightarrow -\om)
\].\no
\eea
\end{widetext}


\begin{thebibliography}{00}
\bibitem{rapp2000}R.~Rapp and J.~Wambach, 
\anp{25}{1}{2000}.

\bibitem{shuryak1978}E.V.~Shuryak, 
\plb{78}{150}{1978};
K.~Kajantie, J.~Kapusta, L.~McLerran and A.~Mekjian,
\prd{34}{2746}{1986};
I. Tserruya, 
\npa{590}{127c}{1995}.

\bibitem{teodorescu2001}
O.~Teodorescu, A.K.~Dutt-Mazumder and C.~Gale,
\prc{63}{034903}{2001}; 
{\bf 66}, {015209} (2002);
C.~Song, P.W.~Xia and C.M.~Ko,
\ibid{52}{408}{1995}. 

\bibitem{gale1991}C.~Gale and J.I.~Kapusta,
\npb{357}{65}{1991}. 

\bibitem{hatsuda1992}T.~Hatsuda and S.H.~Lee,
\prc{46}{R34}{1992};
S.~Leupold, W.~Peters and U.~Mosel,
\npa{628}{311}{1997};
F.~Klingl, N.~Kaiser and W.~Weise,
{\bf A 624}, {527}{ (1997)}. 



 \bibitem{brown1991}G.E.~Brown and M.~Rho, 
\prl{66}{2720}{1991}.

\bibitem{agakichiev1995}G.~Agakichiev {\em et al.}, CERES collaboration, 
\prl{75}{1272}{1995}; 
P.~Wurm, CERES collaboration, 
\npa{590}{103c}{1995}.

\bibitem{ligq1995}Guo-Qiang Li, C.M.~Ko and G.E.~Brown,
\prl{75}{4007}{1995}; \npa{606}{568}{1996}.

\bibitem{kapusta1989}J.I.~Kapusta, {\it Finite
Temperature Field Theory} (Cambridge University Press, Cambridge, England,
1989).

\bibitem{chin1977}S.A.~Chin, 
\ann{108}{301}{1977}.

\bibitem{saito1989}K.~Saito, T.~Maruyama and K.~Soutome,
\prc{40}{407}{1989}; 
K.~Saito and A.~W.~Thomas, 
\ibid{51}{2757}{1995}.

\bibitem{rebhan2001}A.~Rebhan, 
hep-ph/0111341.

\bibitem{eletsky1993}V.L.~Eletsky, J.I.~Kapusta and R.~Venugopalan,
\prd{48}{4398}{1993};
M.~Prakash, R.~Rapp, J.~Wambach and I.~Zahed,
\prc{65}{034906}{2002}.

\bibitem{chanfray2001}G.~Chanfray and M.~Erickson, \epja{16}{291}{2003}.

\bibitem{shiomi1994}H.~Shiomi and T.~Hatsuda, \plb{334}{281}{1994};
 T.~Hatsuda, H.~Shiomi and H.~Kuwabara, \ptp{95}{1009}{1996}.

\bibitem{chen2002}Ji-sheng Chen, Jia-rong Li and Peng-fei Zhuang,
\jhep{11}{014}{2002}.

\bibitem{serot1986}B.D.~Serot and J.D.~Walecka, \anp{16}{1}{1986}.



\bibitem{machleidt1987}R.~Machleidt, K.~Holinde and Ch.~Elster,
\pr{149}{1}{1987};
R. Machleidt, \anp{19}{189}{1989}.

\bibitem{kurasawa1988}H.~Kurasawa and T.~Suzuki, 
\npa{490}{571}{1988}.

\bibitem{jean1994}H.-C.~Jean, J.~Piekarewicz and  A.G.~Williams,
\prc{49}{1981}{1994}.


\bibitem{dutt1997}A.K.~Dutt-Mazumder, B.~Dutta-Roy and A.~Kundu,
\plb{399}{196}{1997};
S.~Sarkar, J.~Alam, P.~Roy, A.K.~Dutt-Mazumder, B.~Dutt-Roy and B.~Sinha,
\npa{634}{206}{1998}.



\bibitem{iwasaki00}Y.~Iwasaki, H.~Kouno, A.~Hasegawa and M.~Nakano,
\ijmpe{9}{459}{2000}.

\bibitem{caillon1993}J.C.~Caillon and  J.~Labarsouque,
\plb{311}{19}{1993}.

\end{thebibliography}
\end{document}